\documentclass[twocolumn,showpacs,preprintnumbers,prd,nofootinbib]{revtex4-1}
\usepackage{latexsym}
\usepackage{graphicx}
\usepackage{amsmath}
\usepackage{amsfonts}
\usepackage{amssymb}

\begin{document}

\title{Study of rotation curves of spiral galaxies with a scalar field dark matter model}



\author{M.A. Rodr\'iguez-Meza}
\affiliation{
Departamento de F\'{\i}sica, Instituto Nacional de
Investigaciones Nucleares, 
Apdo. Postal 18-1027, M\'{e}xico D.F. 11801,
M\'{e}xico. E-mail: marioalberto.rodriguez@inin.gob.mx
}

\begin{abstract}
In this work we study rotation curves of spiral galaxies using a model of dark matter based 
on a scalar-tensor theory of gravity.
We 
show how to estimate
the scalar field dark matter parameters using a sample of observed rotation curves.
\end{abstract}

\maketitle

\section{Introduction}\label{sec:Intro}

Modern cosmological observations, 
like galaxies surveys (SDSS, 2dF), galaxy rotation curves, the Bullet Cluster observation, studies
of clusters of galaxies, surveys of supernovae Ia, the cosmic microwave background radiation, and
the primordial nucleosynthesis
establish that the Universe behaves as dominated
by dark matter (DM) and dark energy. However, the direct evidence for the existence of  these
invisible components remains lacking. Several theories that would modify our understanding
of gravity have been proposed in order to explain the large scale structure formation in the 
Universe
and the galactic dynamics. 

During the last decades there have been several proposals to explain DM, for example: 
Massive Compact Halo Objects
(Machos), 
Weakly Interacting Massive Particles (WIMPs).
Other models propose that
there is no dark matter and use 
general relativity with an appropriate equation of state. 
Or we can use 
scalar fields, minimally or non minimally coupled to the geometry.

In this work we are mainly concern with the study of dark matter (DM) and its influence on the 
rotation curves of spiral galaxies.
Scalar fields are the most natural candidates to model dark matter models.
Our DM model is based on using a scalar field (SF) that
is coupled non-minimally with the Ricci scalar in the Lagrangian that gives Einstein field equations. 

On galactic scales we can test dark matter models using observed rotation curves. 
Nowadays we have lots of compiled samples of galaxy data. The best type of galaxies 
to test dark matter model is the low surface brightness type of galaxies. In this work we
test our dark matter model using a sample of low and high surface brightness galaxies. 

We organize
our work in the following form: In the next section we present the general theory of a typical
scalar-tensor theory (STT), i. e., a theory that generalizes Einstein's general relativity by including
the contribution of a scalar field that couples non-minimally to the metric. 
Next we present our results 
for 
the estimation of the parameters of the scalar field dark matter model.
Finally, 
our conclusions are given in the last section.

\section{General scalar-tensor theory and its Newtonian limit}\label{sec:STT}

We start with the Lagrangian of a general scalar-tensor theory
\begin{equation}\label{EqSTTLagrangian}
{\cal L} = \frac{\sqrt{-g}}{16\pi} \left[ -\phi R + \frac{\omega(\phi)}{\phi}
	(\partial \phi)^2 - V(\phi) \right] + {\cal L}_M(g_{\mu\nu}) \; .
\end{equation}
Here $g_{\mu\nu}$ is the metric,
${\cal L}_M(g_{\mu\nu})$ is the matter Lagrangian and $\omega(\phi)$ and
$V(\phi)$ are arbitrary functions of the scalar field. The fact that we have
a potential term $V(\phi)$ tells us that we are dealing with a massive scalar field.
Also, the first term in the brackets, $\phi R$, is the one that gives the name of
non-minimally coupled scalar field.

When we make the variations of the action, $S=\int d^4x\, {\cal L}$,
with respect to the metric and the scalar field we
obtain the
 Einstein field equations \cite{Faraoni2004}
\begin{eqnarray}\hspace{-1em}
R_{\mu\nu} - \frac{1}{2} g_{\mu\nu} R &=& \frac{1}{\phi}
\left[ 8 \pi T_{\mu\nu} + \frac{1}{2} V g_{\mu\nu}
+ \frac{\omega}{\phi} \partial_\mu \phi \partial_\nu \phi
\right. \nonumber \\
&& \left. -\frac{1}{2} \frac{\omega}{\phi}(\partial \phi)^2 g_{\mu\nu}
+ \phi_{;\mu\nu} - g_{\mu\nu} \, \square \phi \frac{\mbox{}}{\mbox{}}
\right]  , \label{eq:EinsteinEqs}
\end{eqnarray}
for the metric $g_{\mu\nu}$ and for the massive SF $\phi$ we have
\begin{equation}\label{EqSTTPhi}
\square \phi  = \frac{1}{3+2\omega} \left[
	8\pi T -\omega' (\partial \phi)^2 +\phi V' - 2V \right] \, ,
\end{equation}
where $()' \equiv \frac{\partial }{\partial \phi}$. Here $T_{\mu\nu}$ is the energy-momentum
tensor with trace $T$, $\omega(\phi)$ and $V(\phi)$ are in general arbitrary functions that 
govern kinetic and potential contribution of the SF. 
The potential contribution, $V(\phi)$,
provides mass to the SF, denoted here by $m_{SF}$.
%

\subsection{Newtonian limit of a STT}\label{sec:NewtonianLimitSTT}

To study the rotation curves of galaxies we 
need to consider the influence
of SF in the limit of a static STT, and then we need to describe
the theory in its Newtonian approximation, that is, where  gravity 
and the SF are weak (and time independent) and velocities of dark matter
particles are
non-relativistic.  
We expect to have small deviations of
the SF around the background field, defined here as
$\langle \phi \rangle$ and can be understood as the scalar field beyond all matter.
Accordingly we assume that the SF oscillates around the constant background field
\begin{displaymath}
\phi = \langle \phi \rangle + \bar{\phi}
\end{displaymath}
and
\begin{displaymath}
g_{\mu\nu} = \eta_{\mu\nu} + h_{\mu\nu},
\end{displaymath}
where $\eta_{\mu\nu}$ is the Minkowski metric.
Then,  Newtonian approximation
gives \cite{Pimentel1986,Salgado2002,Helbig1991,mar2004}
\begin{eqnarray}
R_{00} = \frac{1}{2} \nabla^2 h_{00} &=& \frac{G_N}{1+\alpha} 4\pi \rho
- \frac{1}{2} \nabla^2 \bar{\phi}  \; ,
\label{pares_eq_h00}\\
  \nabla^2 \bar{\phi} - m_{SF}^2 \bar{\phi} &=& - 8\pi \alpha\rho \; ,
\label{pares_eq_phibar}
\end{eqnarray}
we have set $\langle\phi\rangle=(1+\alpha)/G_N$ 
and $\alpha \equiv 1 / (3 + 2\omega)$.  
In the above expansion we have set
the cosmological constant term equal to zero, since on small galactic
scales its influence should be negligible.  

Note that equation (\ref{pares_eq_h00}) can be
cast as a Poisson equation for $\psi \equiv (1/2) (h_{00} + \bar{\phi}/\langle\phi\rangle)$,
\begin{equation}\label{eq:Psi}
 \nabla^2 \psi = 4\pi \frac{G_N}{1+\alpha}  \rho
\end{equation}
and the New Newtonian potential is given by 
$\Phi_N \equiv (1/2)h_{00}=\psi-(1/2) \bar{\phi}/\langle\phi\rangle$.
Above equation together  with
\begin{equation}\label{eq:phibar}
  \nabla^2 \bar{\phi} - \lambda^{-2} \bar{\phi} = - 8\pi \alpha\rho \; ,
\end{equation}
form a Poisson-Helmholtz equation and gives
\begin{displaymath}
\Phi_N =\psi-\frac{1}{2} \frac{G_N}{1+\alpha} \bar{\phi}
\end{displaymath}
which represents
the Newtonian limit of the STT with arbitrary potential $V(\phi)$ and function
$\omega(\phi)$ that where Taylor expanded around $\langle\phi\rangle$.
The resulting equations are then distinguished by the constants
$G_N$, $\alpha$, and $\lambda=h_P/m_{SF}c$. Here $h_P$ is Planck's constant.

The next step is to find solutions for this new Newtonian potential given 
a density profile, that is, to find the so--called potential--density pairs. 
General solutions to Eqs. (\ref{eq:Psi}) and (\ref{eq:phibar}) 
can be found in terms of the corresponding Green functions,
and the new Newtonian potential is \cite{mar2004,mar2005}
\begin{eqnarray}
\Phi_N  
&=& - \frac{G_N}{1+\alpha} \int d{\bf r}_s
\frac{\rho({\bf r}_s)}{|{\bf r}-{\bf r}_s|} \nonumber \\
&& -\alpha \frac{G_N}{1+\alpha} \int d{\bf r}_s \frac{\rho({\bf r}_s)
{\rm e}^{- |{\bf r}-{\bf r}_s|/\lambda}}
{| {\bf r}-{\bf r}_s|} + \mbox{B.C.} \label{pares_eq_gralPsiN}
\end{eqnarray}
The first term of Eq. (\ref{pares_eq_gralPsiN}), is the
contribution of the usual Newtonian gravitation (without SF), 
while information about the SF is contained in the
second term, that is, arising from the influence function determined by the
modified Helmholtz Green function, where the coupling $\omega$ ($\alpha$) enters
as part of a source factor.

The potential of a single particle of mass $m$ can be easily obtained from
(\ref{pares_eq_gralPsiN}) and is given by
\begin{equation}\label{eq:pointMassPotential}
\Phi_N = -\frac{G_N}{1+\alpha}\frac{m}{r} \left(
1 + \alpha e^{-r/\lambda}
\right)
\end{equation}
For local scales, $r\ll \lambda$, deviations from the Newtonian theory are exponentially
suppressed, and for $r\gg \lambda$ the Newtonian constant diminishes (augments)
to $G_N/(1+\alpha)$ for positive (negative) $\alpha$. This means that equation
(\ref{eq:pointMassPotential}) fulfills all local tests of the Newtonian dynamics, and it is
only constrained by experiments or tests on scales larger than --or of the order of--
$\lambda$, which in our case is of the order of galactic scales. 
In contrast, the potential in the form of equation 
$\Phi_N= -G\frac{m}{r} \left( 1 + \alpha e^{-r/\lambda} \right)$,
with the gravitational
constant defined as usual, does not fulfills the local tests of the Newtonian dynamics
\citep{Fischback1999}.

\subsection{Multipole expansion of the Poisson-Helmholtz equations}\label{sec:Multipole}
The solutions for a spherically symmetric distribution of mass is as follows:
The Poisson's Green function can be expanded in terms of the spherical 
harmonics, $Y_{ln}(\theta,\varphi)$,
\begin{displaymath}
\frac{1}{|{\bf r}-{\bf r}_s|}=4\pi \sum_{l=0}^\infty \sum_{n=-l}^l \frac{1}{2l+1}
\frac{r_<^l}{r_>^{l+1}} Y_{ln}^*(\theta',\varphi') Y_{ln}(\theta,\varphi), 
\end{displaymath}
where $r_<$ is the smaller of $|{\bf r}|$ and $|{\bf r}_s|$, and
$r_>$ is the larger of $|{\bf r}|$ and $|{\bf r}_s|$ and it allows us that
the standard gravitational potential due to a distribution of mass $\rho({\bf r})$,
without considering the boundary
condition, can be written as   \citep{Jackson1975}
\begin{displaymath}
\psi({\bf r}) = \psi^{(i)}+\psi^{(e)}
\end{displaymath}
where $\psi^{(i)}$ ($\psi^{(e)}$) are the internal (external) multipole expansion of 
$\psi$,
\begin{eqnarray}
\psi^{(i)} &=& - \sum_{l=0}^\infty \sum_{n=-l}^l \frac{\sqrt{4\pi}}{2l+1} q_{ln}^{(i)} 
Y_{ln}(\theta,\varphi) r^{l} \, ,  \nonumber \\
\psi^{(e)} &=& - \sum_{l=0}^\infty \sum_{n=-l}^l \frac{\sqrt{4\pi}}{2l+1} q_{ln}^{(e)} 
\frac{Y_{ln}(\theta,\varphi)}{r^{l+1}} \, , \nonumber
\end{eqnarray}

Here, the 
coefficients of the expansions $\psi^{(i)}$ and $\psi^{(e)}$, known as internal and external 
multipoles, respectively, are given by
\begin{eqnarray}\label{imultipoles_psi_eq}
q_{ln}^{(i)} &=&  \sqrt{4\pi} \int_{V(r\le r')} d{\bf r}' \frac{1}{r'^{l+1}} Y_{ln}^*(\theta',\varphi') \rho({\bf r}')  \, , \nonumber \\
q_{ln}^{(e)} &=& \sqrt{4\pi}  \int_{V(r>r')} d{\bf r}' Y_{ln}^*(\theta',\varphi') r'^l \rho({\bf r}')  \, .
\nonumber
\end{eqnarray}

The integrals are done in a region $V$ where $r\le r'$ for the internal multipoles and in a region
$V$ where $r>r'$ for the external multipoles.
They have the property
\begin{eqnarray}
\begin{array}{l}
q_{l(-n)}^{(i)}=(-1)^n (q_{ln}^{(i)})^* \\[.1in]
q_{l(-n)}^{(e)}=(-1)^n (q_{ln}^{(e)})^*  
\end{array}
\end{eqnarray}

We may write expansions above in cartesian coordinates up to the monopole. 
For the internal
multipole expansion we have
\begin{equation}\label{cartesian_ipsi_eq}
\psi^{(i)} = -M^{(i)} 
\, ,
\end{equation}
and its force is
\begin{equation}\label{cartesian_iFpsi_eq}
\mathbf{F}_\psi^{(i)} = 0
\, ,
\end{equation}
where
\begin{equation}
M^{(i)}\equiv \int_{V(r\le r')} d{\bf r}' \frac{1}{r'}  \rho({\bf r}')  \, ,
\end{equation}

For the external multipoles we have
\begin{equation}\label{cartesian_epsi_eq}
\psi^{(e)} = -\frac{M^{(e)}}{r} 
\, ,
\end{equation}
and its force is
\begin{eqnarray}\label{cartesian_eFpsi_eq}
\mathbf{F}_\psi^{(e)} &=& -\frac{M^{(e)}}{r^3}\mathbf{r} 
 \, ,
\end{eqnarray}
where
\begin{equation}
M^{(e)}\equiv \int_{V(r> r')} d{\bf r}'  \rho({\bf r}')  \, ,
\end{equation}

The external monopole have the usual meaning, i.e., $M^{(e)}$ is the mass
of the volume $V(r>r')$.
We may atach to the internal monopole similar meaning, i.e., $M^{(i)}$ is the internal ``mass''
of the volume $V(r\le r')$.

In the case of the scalar field, with the expansion
\begin{eqnarray}
\frac{\exp(-m |{\bf r}-{\bf r}_s|)}{|{\bf r}-{\bf r}_s|}&=&
4\pi m \sum_{l=0}^\infty \sum_{n=-l}^l 
i_l(m r_<) k_l(m r_>) 
\nonumber \\ && \times
Y_{ln}^*(\theta',\varphi') Y_{ln}(\theta,\varphi) \; ,  \nonumber
\end{eqnarray}
the contribution of the scalar field to the Newtonian gravitational potential 
can be written as 
\begin{displaymath}
\bar{\phi}({\bf r}) = \bar{\phi}^{(i)}+\bar{\phi}^{(e)}
\end{displaymath}
where, for simplicity of notation, we are using $m=m_{SF}=h_P/(c \lambda)$ and
\begin{eqnarray}
\frac{1}{2\alpha} \bar{\phi}^{(i)} &=& \sqrt{4\pi}
\sum_{l=0}^\infty \sum_{n=-l}^l  \frac{i_l(m r)}{(m r)^{l}} \,  \bar{q}_{ln}^{(i)} 
r^{l}Y_{ln}(\theta,\varphi) \, , \nonumber \\
\frac{1}{2\alpha} \bar{\phi}^{(e)} &=& \sqrt{4\pi}
\sum_{l=0}^\infty \sum_{n=-l}^l (m r)^{l+1} k_l(m r) \,  \bar{q}_{ln}^{(e)} 
\frac{Y_{ln}(\theta,\varphi)}{r^{l+1}} \, , \nonumber
\end{eqnarray}
$i_l(x)$ and $k_l(x)$ are the modified spherical Bessel functions.

We 
have defined the multipoles for the scalar field as
\begin{eqnarray}\label{imultipoles_phi_eq}
\bar{q}_{ln}^{(i)} &=& \sqrt{4\pi} \int_{V(r\le r')} d{\bf r}' \, 
\frac{Y_{ln}^*(\theta',\varphi')}{r'^{l+1}}\, (m r')^{l+1} k_l(m r') \, \rho({\bf r}') , \nonumber \\
\bar{q}_{ln}^{(e)} &=& \sqrt{4\pi} \int_{V(r>r')} d{\bf r}' \, Y_{ln}^*(\theta',\varphi')\, 
\frac{i_l(m r')}{(m r')^l} r'^l\, \rho({\bf r}') \, . \nonumber
\end{eqnarray}

They, also, have the property
\begin{eqnarray}
\begin{array}{l}
\bar{q}_{l(-n)}^{(i)}=(-1)^n (\bar{q}_{ln}^{(i)})^* \\[.1in]
\bar{q}_{l(-n)}^{(e)}=(-1)^n (\bar{q}_{ln}^{(e)})^*  
\end{array}
\end{eqnarray}

The above expansions of SF contribution to the Newtonian potential can be written
in cartesian coordinates. The internal multipole expansion of the SF contribution, 
up to the monopole is
\begin{eqnarray}
\frac{1}{2\alpha} \bar{\phi}^{(i)} &=& i_0(mr) \bar{M}^{(i)} 
\label{cartesian_iphi_eq}
\end{eqnarray}
and its force is
\begin{eqnarray}
\frac{1}{2\alpha}\mathbf{F}_\phi^{(i)} &=&
-m^2 \frac{i_1(mr)}{mr} \bar{M}^{(i)} \mathbf{r} 
\label{cartesian_iFphi_eq} 
\, ,	
\end{eqnarray}
where
\begin{equation}
\bar{M}^{(i)}\equiv \int_{V(r\le r')} d{\bf r}'  (mr') k_0(mr')
\frac{1}{r'}  \rho({\bf r}')  \, ,
\end{equation}

In the exterior region the SF monopole contribution to the potential is 
\begin{eqnarray}
\frac{1}{2\alpha} \bar{\phi}^{(e)} &=& mr\, k_0(mr) \frac{\bar{M}^{(e)}}{r}
\label{cartesian_ephi_eq}
\end{eqnarray}
and its force is
\begin{eqnarray}
\frac{1}{2\alpha}\mathbf{F}_\phi^{(e)} &=&
(mr)^2 k_1(mr) \frac{\bar{M}^{(e)}}{r^3} \mathbf{r} 
\label{cartesian_eFphi_eq} 
\, ,	
\end{eqnarray}
where
\begin{equation}
\bar{M}^{(e)}\equiv \int_{V(r> r')} d{\bf r}'  i_0(mr') \rho({\bf r}')  \, ,
\end{equation}

In the limit when $m\rightarrow 0$ we recover the standard Newtonian potential and force
expressions.


\section{Results}\label{sec:Results}


In this section we will show how to obtain values of parameters of the model, i.e., $\alpha$
and $\lambda$ using observed rotation curves of galaxies.

Our SF galaxy model will be as follows. A test particle will move under the action of the 
potential
\begin{equation}
\phi (r) = \phi_D (r) + \phi_{NDM} (r) + \phi_{SF} (r)
\end{equation}
where $\phi_{D} (r)$ is the potential due to the disk mass density profiles of gas and stars, 
$\rho_D$.
$\phi_{NDM} (r)$ is the Newtonian
potential due to a DM mass distribution, and $\phi_{SF} (r)$ is the potential due to the
SF contribution. They satisfy the equations:
\begin{equation}
\nabla^2 \phi (r) = 4\pi G \rho_D (r)
\end{equation}
\begin{equation}
\nabla^2 \phi_{NDM} (r) = 4\pi \frac{G}{1+\alpha} \rho_{DM}(r)
\end{equation}
and
\begin{equation}
\nabla^2 \phi_{SF} (r) - \lambda^{-2} \phi_{SF}(r) = -8\pi \alpha  \rho_{DM}(r) 
\end{equation}

The rotational velocity is obtained using the expression $V=r | d \phi /d r |$.
For an exponential disk
we have the simple approximation\cite{Freeman:1970}
\begin{eqnarray}
V^2_D(r) &=&  \frac{G M_D}{2 R_D} \left(\frac{r}{R_D}\right)^2 \times \\
&&
\left(
I_0(r/2R_D) K_0(r/2R_D)
-I_1(r/2R_D) K_1(r/2R_D)
\right)
\nonumber
\end{eqnarray}
where $I_0(y)$, $I_1(y)$, $K_0(y)$, and $K_1(y)$, are the modified Bessel functions.
$M_D$ is the total mass of the disk and $R_D$ is its scale length.

We consider that the DM mass density profile is given by the pseudo isothermic profile,
\begin{equation}
\rho_{PISO}(r) = \frac{\rho_s}{1+(r/r_s)^2} \, ,
\end{equation}
where $\rho_s$ is the central density of the halo and $r_s$ is its radius. This density profile 
has the force contribution given by
\begin{eqnarray}
\left(
\frac{G_N}{1+\alpha} \frac{M_0}{a_0^2}
\right)^{-1}
F_N  
&=& - \frac{\mathbf{r}}{r^3} m(r) \\
&& - \frac{\mathbf{r}}{r^3} \alpha
\left[
\left(
\frac{r}{\lambda}
\right)^2 k_1(r/\lambda) p(r) 
\right. \nonumber \\ && \left.
-  \left(
\frac{r}{\lambda}
\right)^2 \lambda
i_1(r/\lambda) q(r)
\right] \nonumber
\end{eqnarray}
with the auxiliary functions,
\begin{eqnarray}
m(r) &=& 4\pi \int_0^r dr' r'^2 \rho(r') \, ,\\
p(r) &=& 4\pi \int_0^r dr' r'^2 i_0(r/\lambda) \rho(r') \, ,\\
q(r) &=& 4\pi \int_r^\infty dr' r' (r'/\lambda)k_0(r'/\lambda) \rho(r') \, .
\end{eqnarray}
If we define the function
\begin{eqnarray}
f(x) &=& 
\left(
\cosh[r/\lambda] - \frac{\sinh[r/\lambda]}{[r/\lambda]}
\right)
q(r)  \nonumber \\
&&
- p(r) \left( \frac{1}{\lambda} + \frac{1}{r}\right)
\exp[-r/\lambda]
\end{eqnarray}
Then,
\begin{eqnarray}
V^2_{SF}(r) = V^2_0 
\left| 
\frac{1}{1+\alpha} 
\left(
\frac{m(r)}{r} - \alpha f(r)
\right)
\right|
\end{eqnarray}
and the rotation curve is obtained from
\begin{eqnarray}
{
V^2(r)=V^2_{D}(r)+V^2_{SF}(r)
}
\end{eqnarray}
We $\chi^2$ best fit the observations with
\begin{equation}
\chi^2 = \sum_{i=1}^N 
\left(
\frac{V_i - V(r_i)}{\sigma_i}
\right)^2
\end{equation}
Free parameter are: $r_s$, $\rho_s$, $\alpha$, and $\lambda$.
Units we are using are such that $a=1$ kpc $=1$, $V_0=1$ km/s  $=1$ and $G=1$.

In table \ref{rc_table} we show properties and parameters of the analyzed sample of 
galaxies. 
This sample include low and high surface brightness:
UGC 4325 is a late-type dwarf galaxy at a distance of 10.1 Mpc; has
an absolute $B$-band magnitude of $M_B=-17.2$ \cite{Kuzio2008}. 
DDO 47 is a nearby dwarf galaxy of type IBm \cite{Gentile2005}.
NGC 3109 is a nearby Magellanic-type galaxy with a $M_B=-16.35$ \cite{Valenzuela2007}.
ESO 116-G12  is a SBm type galaxy with magnitud $M_I = -20.0$ \cite{Gentile2004}.
Galaxy NGC 7339 is a SABbc type galaxy with magnitud $M_I = -20.6$ \cite{Gentile2004}.
NGC 6822 is a nearby Local Group member galaxy with a $B$-magnitude of 
$M_B=-15.8$ \cite{Valenzuela2007}.
Andromeda galaxy M31  is a very large local group spiral galaxy, 
SAb type, at a distance of 0.78 Mpc \cite{Corbelli2007}.
UGC 8017 is a Sab type galaxy at a distance of 102.7 Mpc \cite{Vogt2004}.
UGC 11455 is a Sc type galaxy at a distance of 75.4 Mpc \cite{Vogt2004}.
UGC 10981 is a Sbc type galaxy at a distance of 155 Mpc  \cite{Vogt2004}.

We have divided the sample in two groups. Group A is composed of galaxies:
UGC 4325, DDO 47, NGC 3109, ESO 116-G12, NGC 7339. Group B is composed of
galaxies: NGC 6822, M31, UGC 8017, UGC 11455 and UGC 10981.

\begin{table}
\begin{tabular}{llccc}
\hline \hline
Galaxy &Type&L ($10^9$ L$_\odot$) & $R_D$ (kpc) & $M_D$ ($10^9$ M$_\odot)$ 
\\ \hline \hline
Group A:  & & & & \\
UGC 4325	& Sm && 1.6 	& $0.8846 $ 
\\
DDO 47 		&IBm&0.1& 0.5 	& 0.01 
\\
NGC 3109 	&SBsm&&  1.2	& 0.11 
\\
ESO 116-G12	&SBm&4.6& 1.7 	&  2.1  
\\
NGC 7339	&SABbc&7.3& 1.5 	&  22 
\\
\hline
Group B:  & & & & \\
NGC 6822 	&&0.1& 0.5 	&  0.01 
\\
M31		 	&SAb&20& 4.5 	&  126 
\\
UGC 8017	&Sab&40&  2.1	& 9.1 
\\
UGC 11455 	&Sc&45&  5.3	& 74 
\\
UGC 10981  	&Sbc&120&  5.4	& 115 
\\ \hline \hline
\end{tabular}
\caption{Properties and parameters of the analyzed 
sample. From left to right, the columns read: 
name of the galaxy;
Hubble clasification;
Luminosity;
disk scale length in kpc; 
disk mass in $10^9 M_\odot$.
}\label{rc_table}
\end{table}

\begin{figure}
\begin{center}
\includegraphics[width=2.5in]{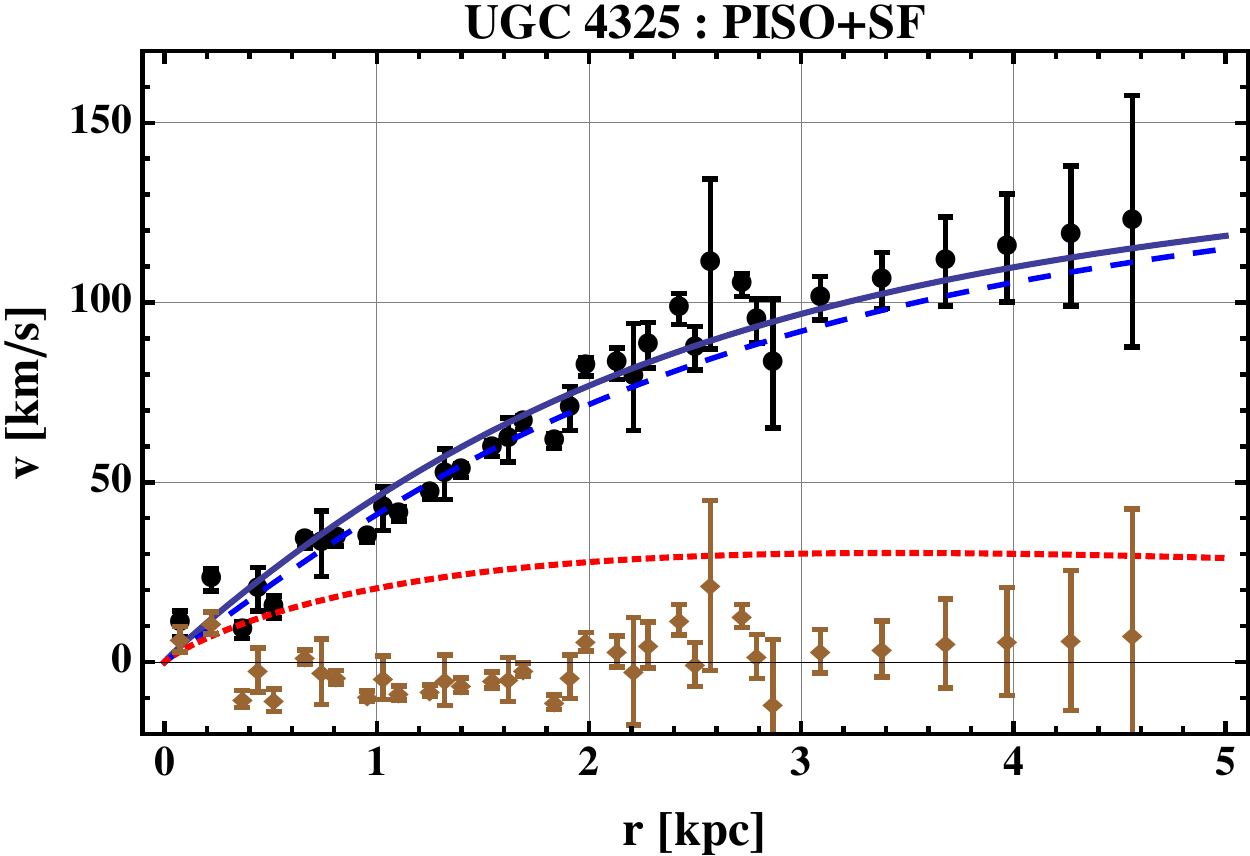} \\ [0.5\baselineskip]
\includegraphics[width=2.5in]{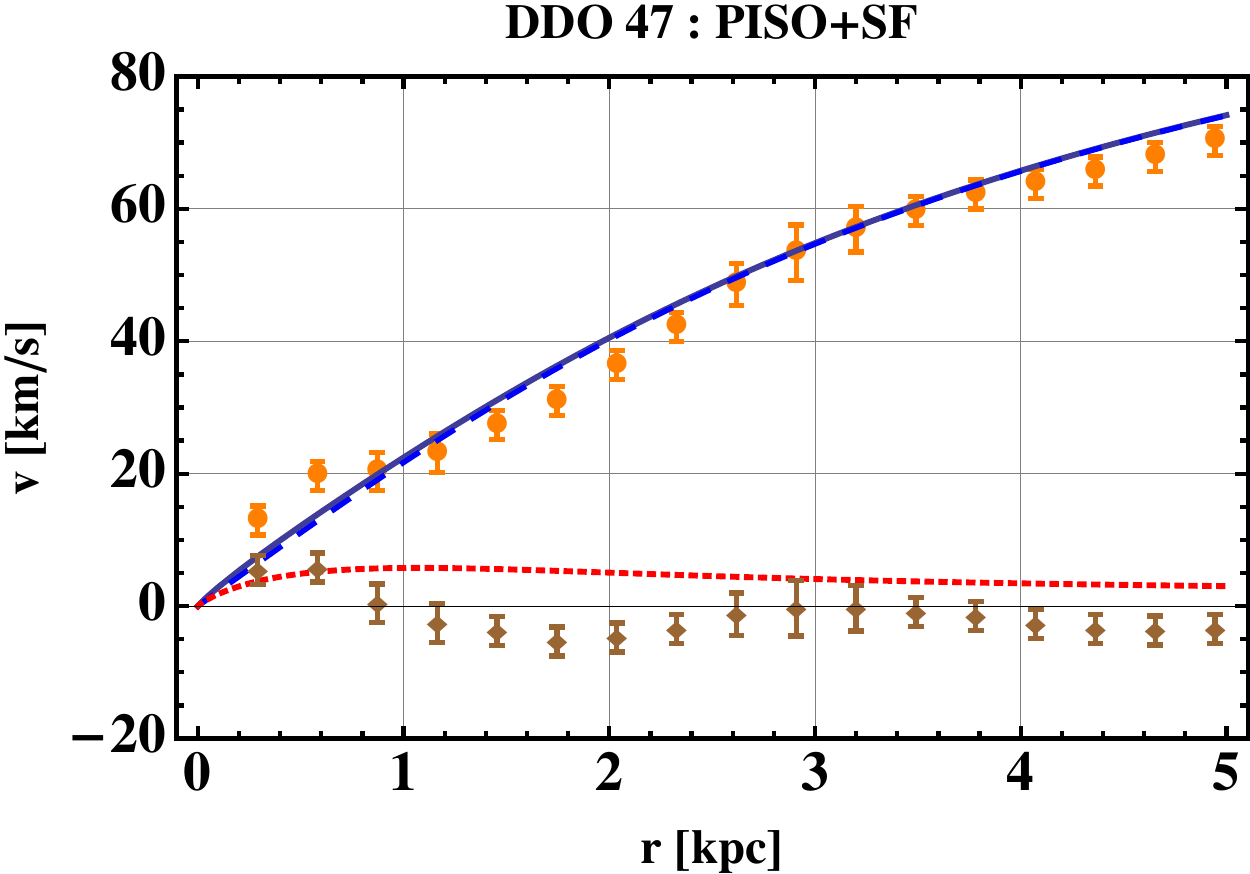} \\  [0.5\baselineskip]
\includegraphics[width=2.5in]{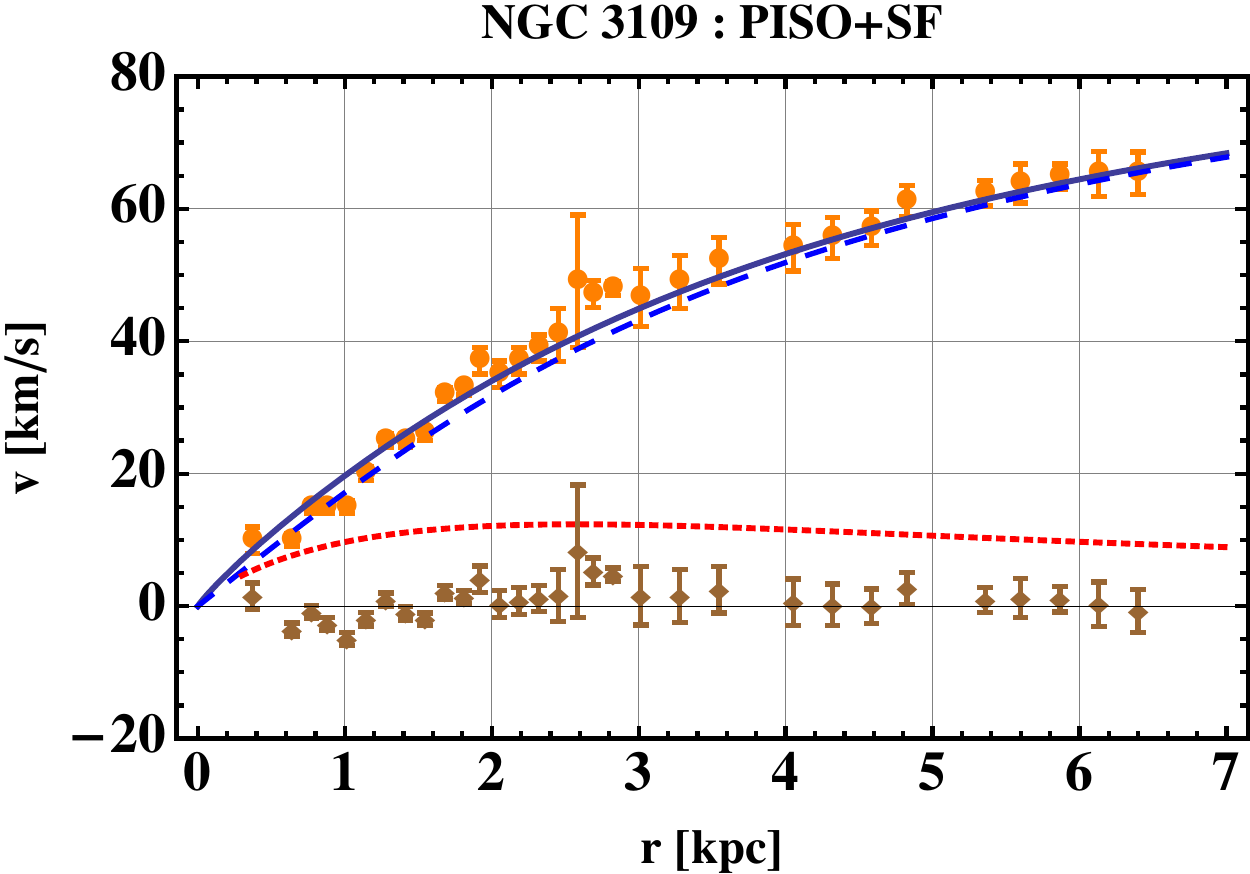} \\ [0.5\baselineskip]
\includegraphics[width=2.5in]{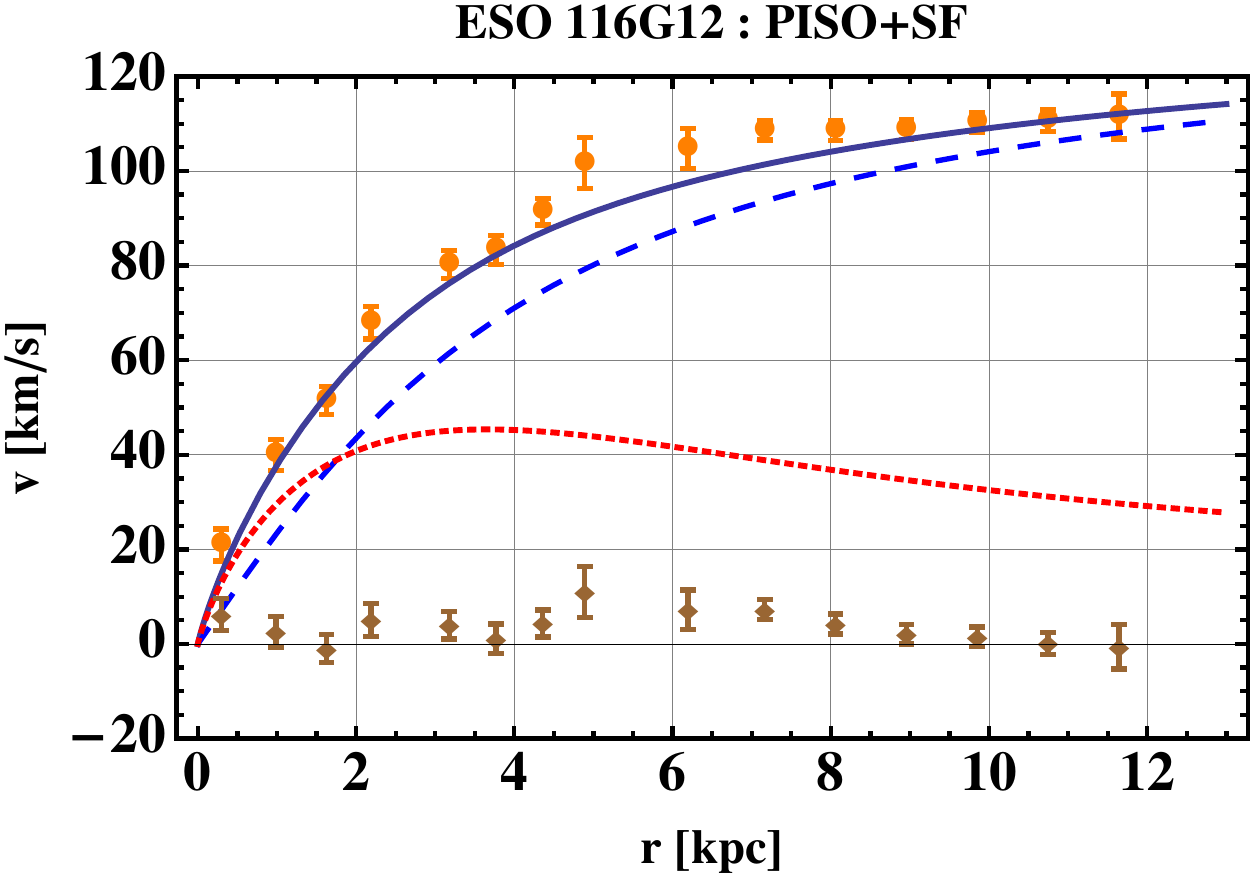} \\ [0.5\baselineskip]
\includegraphics[width=2.5in]{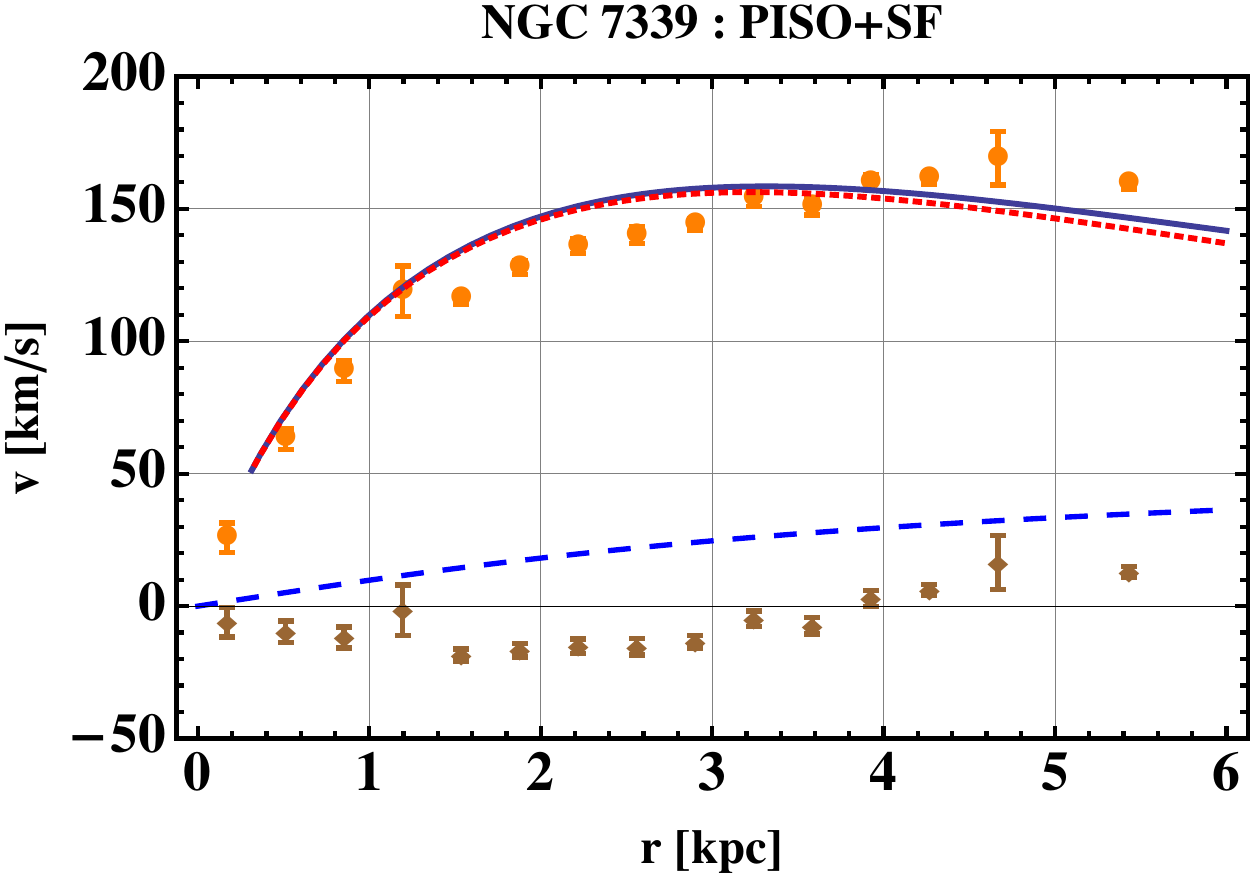}
\end{center}
\caption{Group A of analyzed galaxies: UGC 4325, DDO 47, NGC 3109,
 ESO 116-G12, NGC 7339.}
\end{figure}

\begin{figure}
\begin{center}
\includegraphics[width=2.5in]{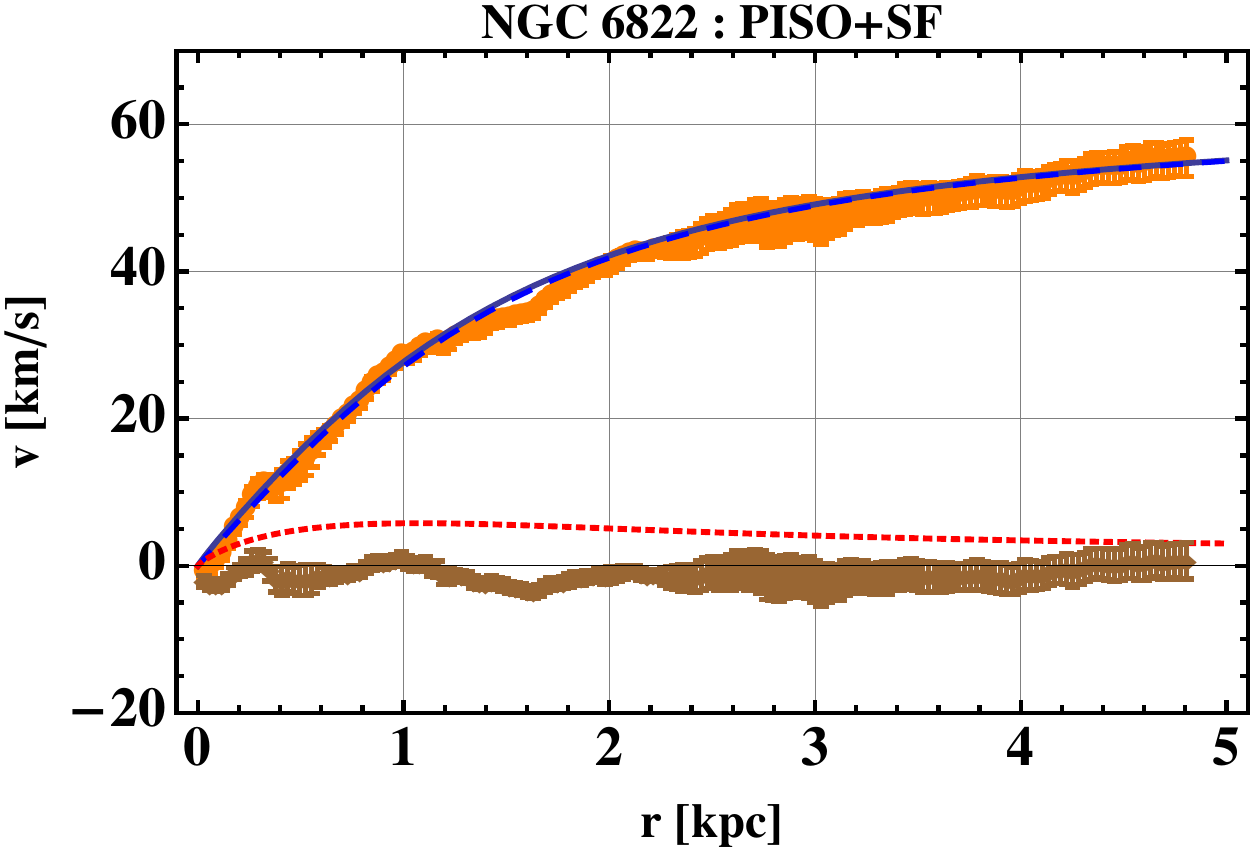}  \\ [0.5\baselineskip]
\includegraphics[width=2.5in]{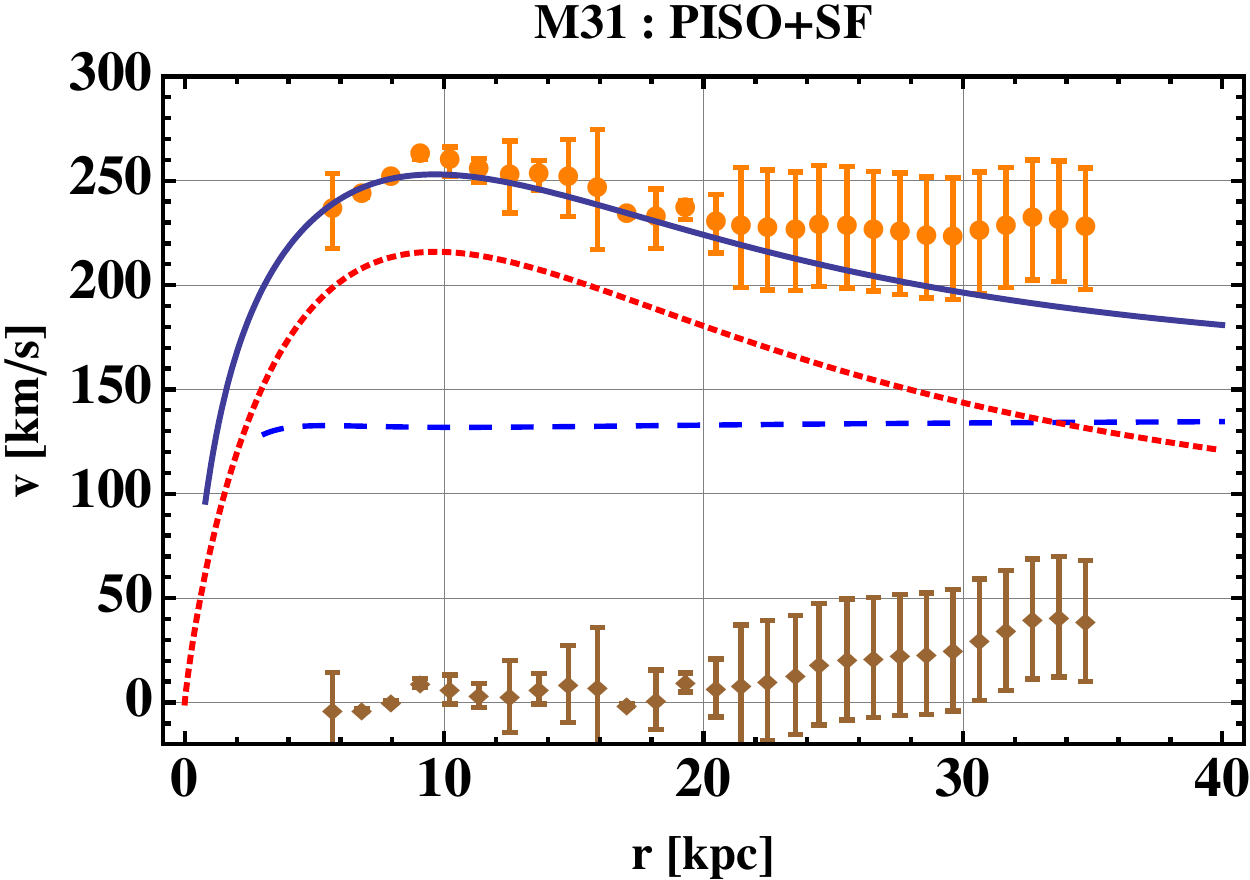} \\ [0.5\baselineskip]
\includegraphics[width=2.5in]{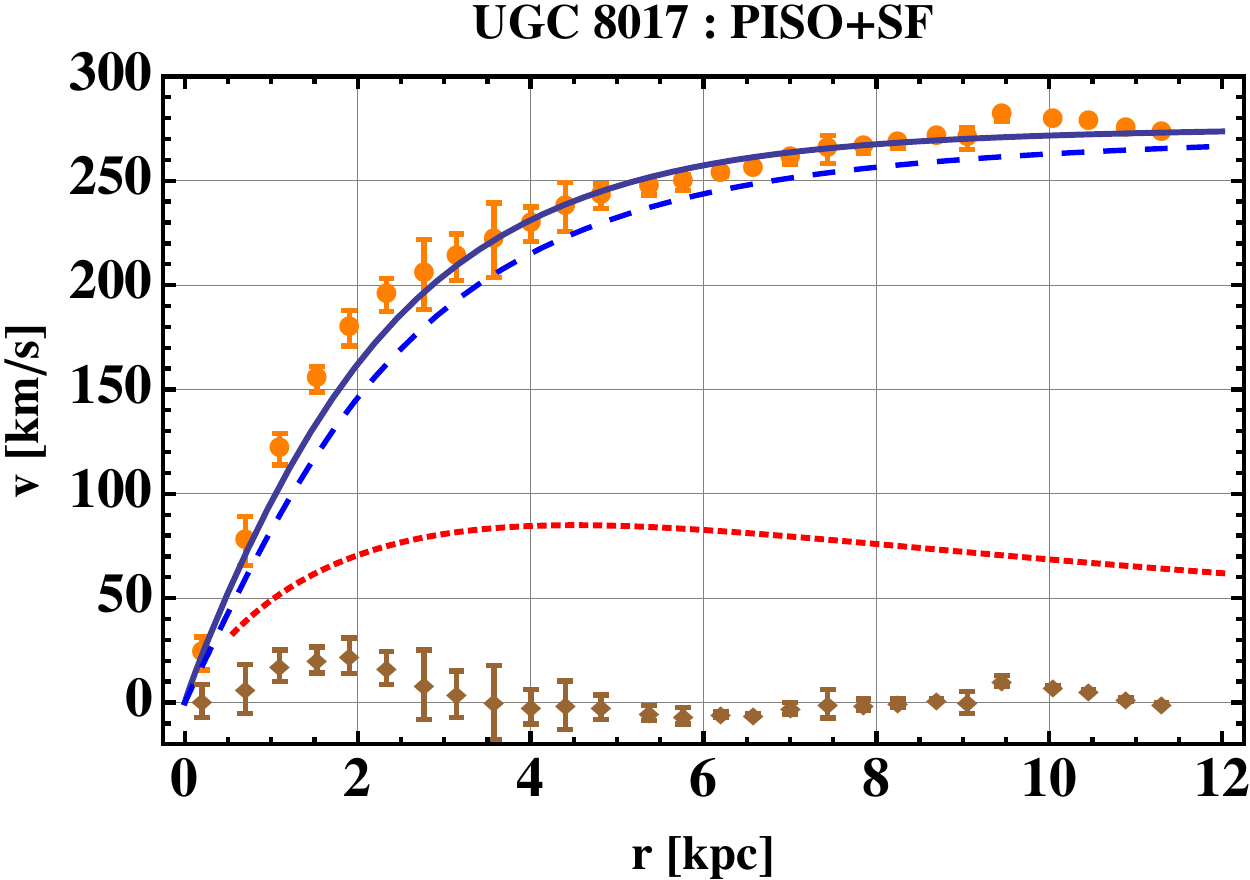} \\ [0.5\baselineskip]
\includegraphics[width=2.5in]{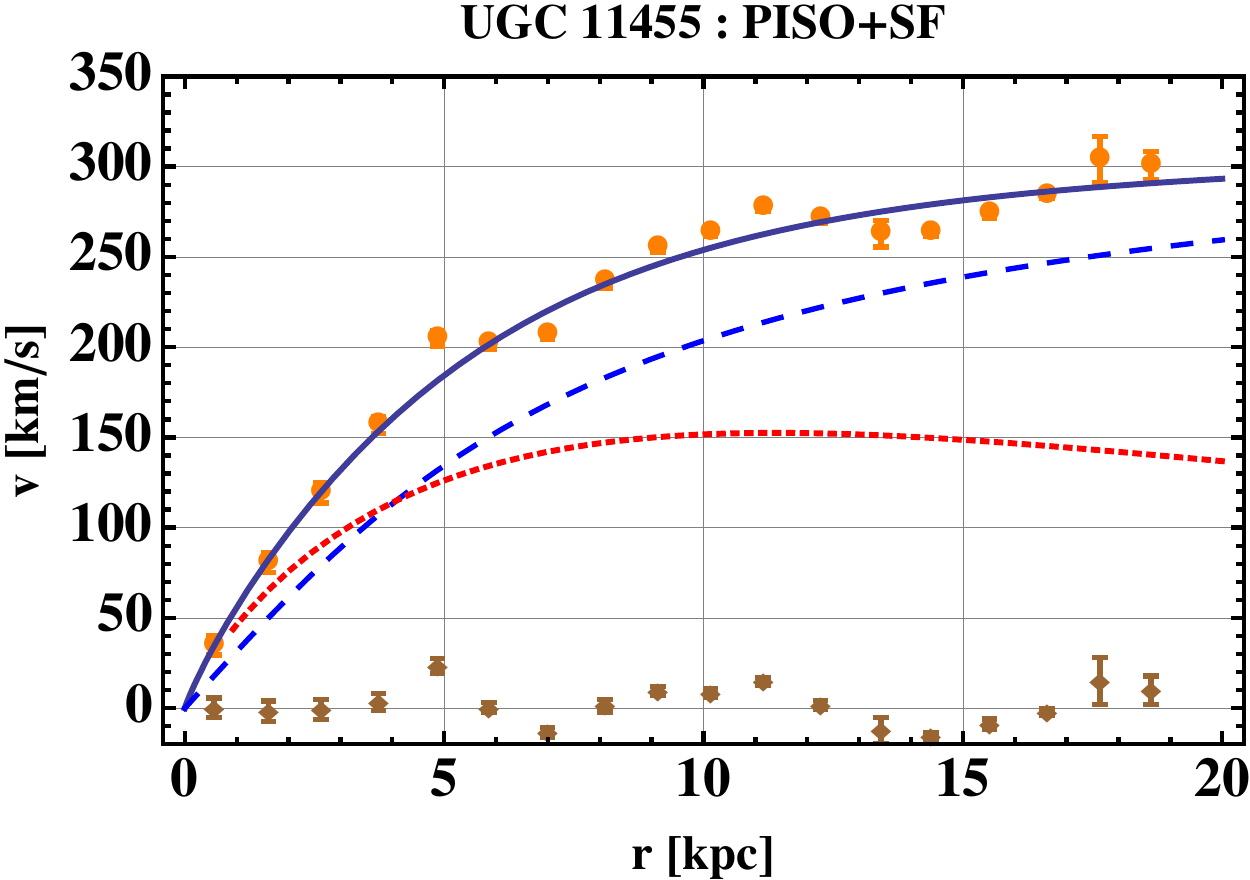}  \\ [0.5\baselineskip]
\includegraphics[width=2.5in]{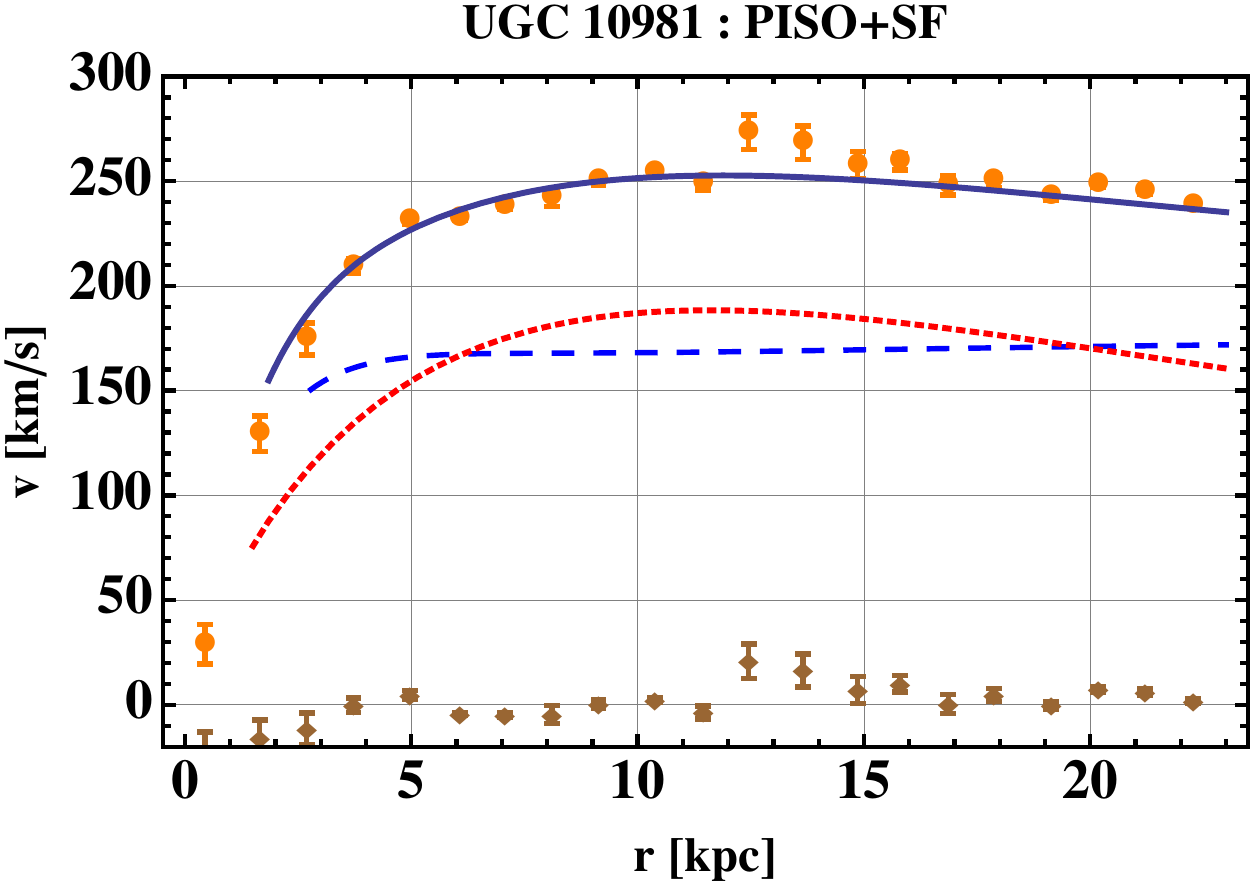}
\end{center}
\caption{Group B of analyzed galaxies:  NGC 6822, M31, UGC 8017, UGC 11455, 
UGC 10981.}
\end{figure}

\begin{table}
\begin{tabular}{lccccr}
\hline \hline
Galaxy & $\rho_s$ ($\rho_0$) & $r_s$ (kpc) & $\lambda$ (kpc) & $\alpha$ & $\chi^2_{red}$ \\ \hline \hline
Group A:  & & & & & \\
UGC 4325	& 3492.04 & 1.95 & 0.691 & -0.477 & 7.20	 \\
DDO 47 		& 1281.99 & 2.983 & 1.976 & -0.259 & 3.35 	\\
NGC 3109 	& 800.66 & 2.98 & 1.98 & -0.26 & 3.85 \\
ESO 116-G12	& 1500 & 2.983 & 1.976 & -0.259 & 3.49	\\
NGC 7339	& 260.66 & 2.98 & 1.98 & -0.26 & 26.89	\\
\hline
Group B:  & & & & & \\
NGC 6822 	& 3501.96 & 1.46 & 1.09 & 0.67 & 2.21	\\
M31		 	& 40001 & 0.98 & 1.79 & 1.05 & 3.50 	\\
UGC 8017	& 26000 & 2.48 & 3.09 & 0.84 & 19.56	\\
UGC 11455 	& 4000.7 & 6.30 & 1.39 & 0.42 & 16.71	\\
UGC 10981  	& 35001 & 1.57 & 1.69 & 1.69 & 10.45	
\\ \hline \hline
\end{tabular}
\caption{Properties and parameters of the analyzed 
sample. From left to right, the columns read: 
name of the galaxy;
central density in units of $\rho_0=18491.4$ solar masses per kpc$^3$;
central radius in kpc;
scalar field scale length in kpc;
scalar field strength;
the $\chi^2_{red}$ value.
}\label{rc_table_fits}
\end{table}

In table \ref{rc_table_fits} we show the fitting results. We observe that for the galaxies
in group A (UGC 4325, DDO 47, NGC 3109, ESO 116G12 and NGC 7339) 
the parameter $\alpha$
of the scalar field dark matter model is negative and $\alpha = -0.303$ on the average. 
For this group the
average value of scale length of the scalar field is $\lambda=1.72$ kpc.
For all 
the rest of galaxies, group B, 
 $\alpha$ is positive with an average value $\alpha = 0.934$. In
this case all galaxies, except NGC 6822, have a high luminosity. For this group the
average value of scale length of the scalar field is $\lambda=1.81$ kpc.

\section{Conclusions}\label{sec:Conclusions}

We have used a general, static STT, that is compatible with local observations by the
appropriate definition of the background field constant, i. e., 
$\langle \phi\rangle = (1+\alpha)/G_N$, to study rotation curves of spiral galaxies.

It is important to note that particles in our model are gravitating particles and that
the SF acts as a mechanism that modifies gravity. The effective mass of the SF
 ($m_{SF}=1/\lambda$) only sets an interaction length scale for the Yukawa term.

We have 
estimated the parameters of the scalar field dark matter model by minimizing
the appropriate $\chi^2$ for two samples of observed rotation curves. 
For the galaxy group A, i.e., the dwarf and low surface brightness 
galaxies, the strength parameter has a negative value, and on the average 
$\alpha = -0.303$ and $\lambda=1.72$ kpc. 
Whereas, for group B, i.e., the high surface brightness galaxies,
$\alpha = 0.934$ and $\lambda=1.81$ kpc. We have found, in general, also higher 
values of
$\chi^2_{red}$ for the group B of galaxies. This should be consistent with the general
believe that low surface brightness galaxies are dominated by dark matter whereas
high surface brightness galaxies are not.



\bibliographystyle{aipproc}   
\bibliographystyle{aipprocl} 

\IfFileExists{\jobname.bbl}{}
 {\typeout{}
  \typeout{******************************************}
  \typeout{** Please run "bibtex \jobname" to optain}
  \typeout{** the bibliography and then re-run LaTeX}
  \typeout{** twice to fix the references!}
  \typeout{******************************************}
  \typeout{}
 }

\end{document}